\documentclass[aps,prb,twocolumn,preprintnumbers,amsmath,amssymb]{revtex4}
\usepackage{graphicx}
\begin{document}
\title{Thermal expansion and magneto-volume studies of the itinerant helical magnet MnSi}

\author{A.E. Petrova}
%\affiliation{Institute for High Pressure Physics of RAS, Troitsk, Russia}
\author{S.M. Stishov}
\email{sergei@hppi.troitsk.ru}
\affiliation{Institute for High Pressure Physics of RAS, Troitsk, Moscow, Russia}

\begin{abstract}
Thermal expansion and forced magnetostriction of MnSi were measured as a function of temperature down to 5 K and magnetic field to 3 T. The small length (volume) discontinuity at the magnetic phase transition in MnSi decreases with application of magnetic field to a value $\Delta L/L \sim 10^{-7}$, and then suddenly the discontinuity seemingly jumps to zero. Thermal expansivity peaks strongly deteriorate with magnetic fields. No specific features identifying a tricritical point were observed.  We propose that the Frenkel concept of heterophase fluctuations may be relevant in the current case. Therefore, we suggest that the magnetic phase transition in MnSi always remains first order at any temperature and magnetic field, but the transition is progressively smoothed by heterophase fluctuations. These results  question the applicability of a model of a fluctuation-induced first order phase transition for MnSi. Probably a model of coupling of an order parameter with other degrees of freedom is more appropriate.
\end{abstract}
\maketitle
%\pacs{75.30.Kz, 72.15.Eb}

\section{Introduction}
The magnetic phase diagram of the itinerant helical magnet MnSi has been studied for over 40 years. From these studies, a magneto-ordered phase in MnSi was identified  as having a helical spin structure~\cite{1}, and a so-called A-phase was discovered~\cite{2} that later was recognized as a spin-skyrmion phase~\cite{3}. Heat capacity, thermal expansion (see for instance~\cite{4,5,6,7} and the references therein) and elastic properties~\cite{8,9} measured across the phase-transition line at various magnetic fields led to the identification of the helical-paramagnetic phase transition at zero magnetic field as a fluctuation-induced weak first order transition~\cite{10,11}. Though the volume and entropy change at the phase transition are so small ($\Delta V/V\approx10^{-6}$, $\Delta S/R\approx 10^{-4}$)~\cite{6,7} that this transition should be considered as "extremely weak first order phase transition".  Indeed, for instance, a volume change at the weakest first order phase transitions in well-known examples, like ferromagnetic transition in  MnAs, the dielectric order-disorder transition in NH$_4$Cl, and ferroelectric transition in KDP is $\Delta V/V \approx 1.8\cdot 10^{-2}$, $1.2\cdot 10^{-3}$, $\sim 10^{-4}$,  respectively (see Refs.~\cite{12,13,14}), which is at least two order of magnitudes larger than  $\Delta V/V$ at the phase transition in MnSi.

One might expect magnetic fluctuations to be suppressed with the application of magnetic fields, which could result in a change of character of the fluctuation-induced first order transition. In this case, the first order phase transition becomes second order at some point in $T$, $H$ space, producing a tricritical point. In the mean field approximation, heat capacity, thermal expansion coefficient and compressibility diverge at the tricritical point as a power law with a temperature exponent $\alpha=0.5$ (see~\cite{15}). The fluctuation correction should be small in this case because the upper critical dimension $D^+=3$ at a tricritical point; whereas, at an ordinary second order transition $D^+=4$ (Ref.~\cite{16}). On the basis of a change of form of heat capacity at the phase transition,  a recent paper~\cite{17},  argued for the discovery of a tricritical point on the $T-H$ transition line in MnSi;  however, there was no tendency for the heat capacity to diverge.   Moreover, ultrasound studies have revealed a striking increase of elastic moduli in the $T-H$ region corresponding to the proposed tricritical point where the elastic moduli should  decrease or the elastic compliance should increase~\cite{9}. With this unresolved issue, it is appropriate to take a closer look at the situation.

To this end, thermal expansion and forced magnetostriction of MnSi were measured as a function of temperature and magnetic field. Though similar measurements were reported earlier in Refs.~\cite{4,5,6,7}, the current measurements were performed in more detail with an emphasis on exploring the disputed region. The main surprising result of this study is a decrease of the extremely small length (volume) discontinuity at the magnetic phase transition in MnSi with application of magnetic field to a value $\Delta L/L \sim 10^{-7}$ and the apparent sudden jump of the discontinuity to zero.

Because no specific features identifying a tricritical point were observed in these experiments,  the field-dependent evolution of the volume discontinuity raises the question of whether extremely weak first order phase transitions with immeasurably small volume and entropy discontinuities may exist~\cite{15}. In our particular case the answer is negative. Instead, heterophase fluctuations smooth the transition to mimic a second order phase transition.

\begin{figure}[htb]
\includegraphics[width=80mm]{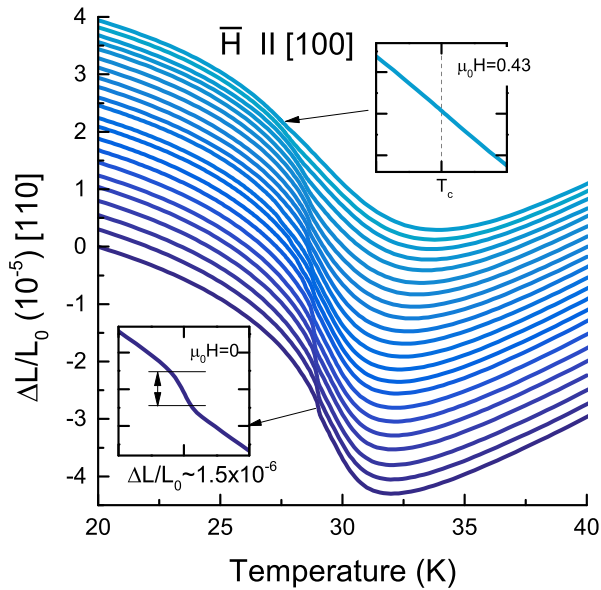}
\caption{\label{fig1} (Color online) Linear thermal expansion of MnSi along [110] as a function of temperature and magnetic field. The magnetic field is directed along [100] (data shown with offsets for better viewing), with field increasing from zero (bottom curve) to 0.43 $T$ (top curve). Insets show enlarged views of corresponding curves near the phase-transition region. Temperature scale division is 0.3 K for both insets,  $\Delta L/L_0$ scale division is $1 \cdot 10^{-6}$ and $2 \cdot 10^{-6}$ for upper and lower insets respectively.}
\end{figure}

\begin{figure}[htb]
\includegraphics[width=80mm]{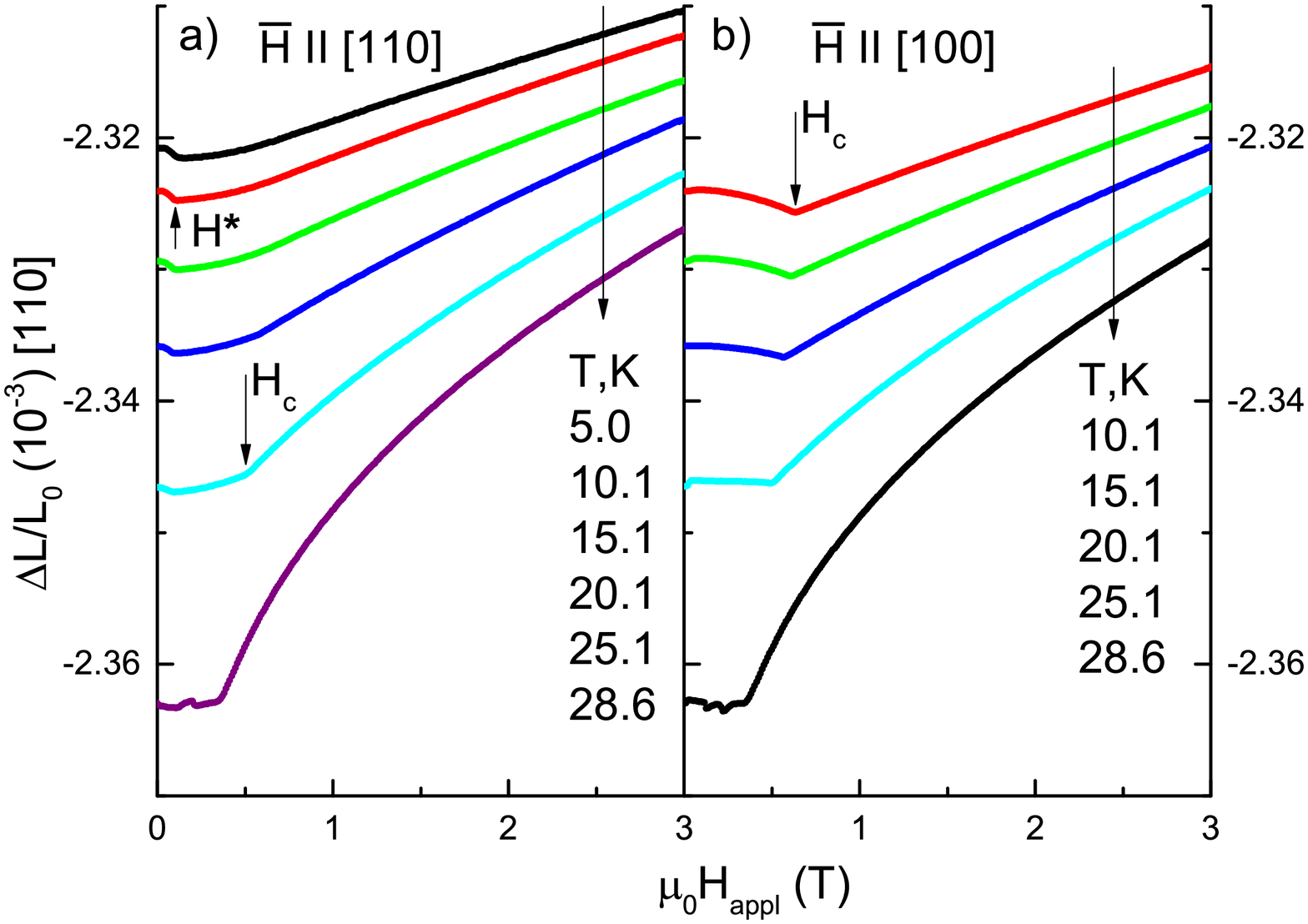}
\caption{\label{fig2} (Color online) Longitudinal (a) and transvers magnetostriction (b) of MnSi along [110]. A low field anomaly $H^\ast$ (seen in panel (a) and in Fig.~\ref{fig3}) corresponds to the helical-conical transition. A change of slope at $H_c$ is the result of a transformation from the conical phase to a field-induced ferromagnetic phase. This effect becomes less pronounced at low temperatures in the longitudinal configuration (see also~\cite{5}).}
 \end{figure}

\section{Experimental}
\begin{figure}[htb]
\includegraphics[width=80mm]{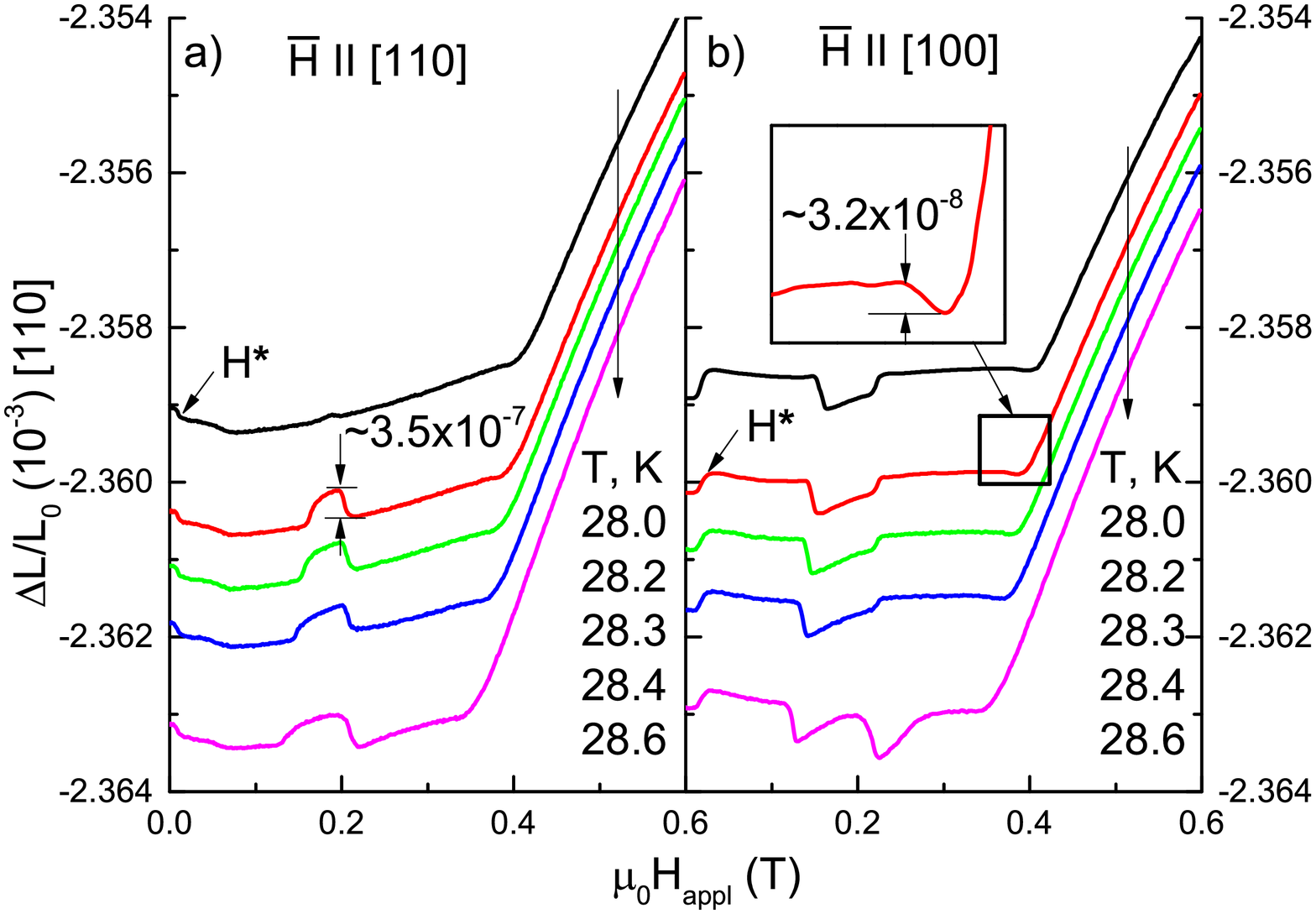}
\caption{\label{fig3} (Color online) Detailed view of longitudinal (a) and transvers magnetostriction (b) of MnSi along [110] in a restricted range of temperatures and magnetic fields.   The low field anomaly at $H^\ast$ corresponds to a helical-conical transition. A higher field change of slope occurs due to a transition of the conical phase to a field-induced ferromagnet state. This transition is associated with tiny dips in $\Delta L/L_0$  that can be seen in (b). These dips can be viewed as continuations of the smoothed first order phase transitions observed at lower magnetic fields. 'Hills' (a) and 'canyons' (b) in the middle characterize the skyrmion phase.}
\end{figure}

A Quantum Design Physical Property Measurement System (PPMS) and a capacity dilatometer~\cite{18} with a resolution of $10^{-8}$~mm were employed in these experiment. A single crystal of MnSi $\sim 3$~mm in diameter and $\sim 2.8$ in height, grown by the Bridgman technique, was used. The dilatometer holder permitted measurements of the sample strains both in "longitudinal" and "transverse" orientations, that is, when the magnetic field was parallel or perpendicular to a direction of the length measurement. In the current experiment, the sample length was always measured in the [110] direction, whereas magnetic field was directed along [110] or [100]. Quite a number of experimental runs determined the thermal expansion at constant magnetic fields and the forced magnetostriction at constant temperatures. Important selected results are shown in Figs.~\ref{fig1}--\ref{fig3}.

The relative change in linear thermal expansion, $\Delta L/L_0$  with $L_0$ being the  sample length at 300~K, along [110] with magnetic fields along [100] is displayed in Fig.~\ref{fig1}. An increased density of the data points around the phase transition region causes a somewhat darker area in Fig.~\ref{fig1}.  The insets in Fig.~\ref{fig1} plot thermal expansion curves in the phase transition area for zero and 0.43 T fields. The thermal expansion in the same direction but with magnetic fields along [110] was also measured.  Differences in these two sets of data are essentially indistinguishable. Note that earlier thermal expansion measurements on MnSi were published exclusively in the form of the thermal expansivity $1/L(dL/dT)$~\cite{4,5,6,7}.

Fig.~\ref{fig2} illustrates behavior of the longitudinal (a) and transverse forced magnetostriction (b) at different temperatures and in magnetic fields to 3 T. The magnetostriction in a narrow range of temperatures near the phase boundary and for small applied magnetic fields, Fig.~\ref{fig3}, reveals a first order phase transition on the boundary of the A-phase.

\begin{figure}[htb]
\includegraphics[width=80mm]{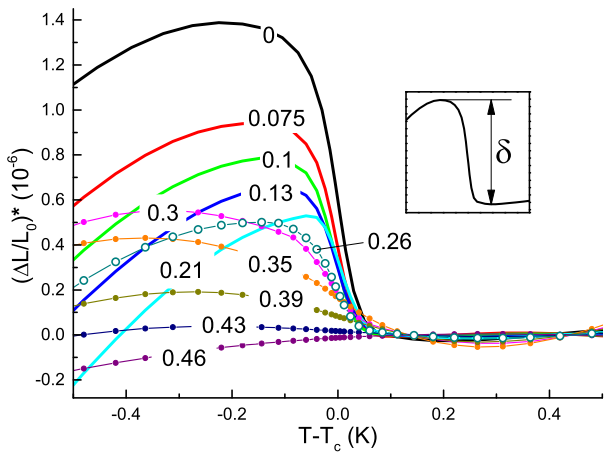}
\caption{\label{fig4} (Color online) Dependence of $\Delta L/L_0 (T)$, after subtracting a linear function that describes $\Delta L/L_0$ in the paramagnetic phase of MnSi in the vicinity of the phase transition (see text).  Numbers on the plot are values of applied magnetic field in Tesla. At $\sim 0.26$ Tesla the form of the thermal expansion curves suddenly changes.  Quasi-discontinuities are replaced by increasingly smoother anomalies. Inset illustrates how the discontinuity was determined.}
\end{figure}
 
\begin{figure}[htb]
\includegraphics[width=80mm]{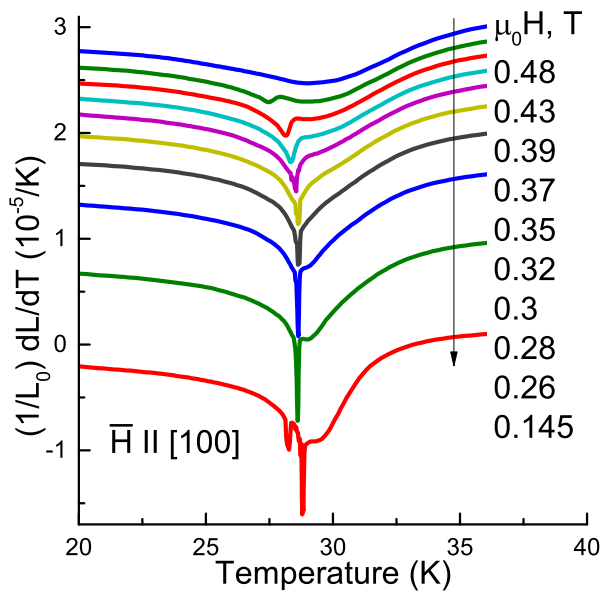}
\caption{\label{fig5} (Color online) Temperature dependence of linear thermal expansivity $1/L(dL/dT)$  along [110] for the applied magnetic field along [100]. Deterioration of the peaks with magnetic field can be clearly seen. A peak in thermal expansivity, corresponding to the skyrmion phase, is apparent at 0.145~T.
}
\end{figure}

\begin{figure}[htb]
\includegraphics[width=80mm]{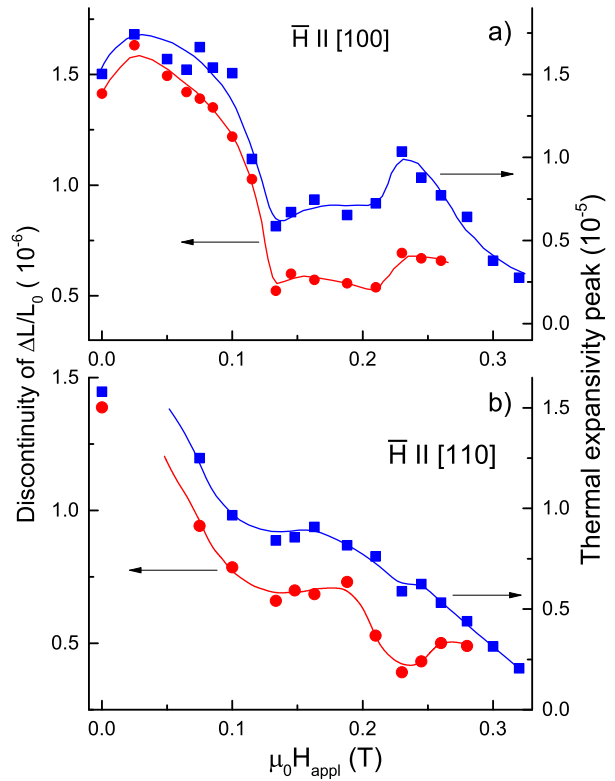}
\caption{\label{fig6} (Color online) Dependence on magnetic field of the quasi-discontinuity of sample length and height of peaks of expansivity  at the phase transition. The discontinuity curves are terminated at 0.26 T, where the discontinuities virtually disappeared.   The expansivity peaks can be still observed at least to 0.43 T.  A plateau in the range $\sim 0.12-–0.2$~T  (a) and a local maximum at $\sim 0.14-–0.23$~T (b) are related to the existence of the skyrmion A-phase (see Fig.~\ref{fig3})}. 
\end{figure}

\section{Discussion}
The small length discontinuities in thermal expansion curves,   manifesting a first order phase transition in MnSi, are situated on an otherwise very steep slope (Fig.~\ref{fig1}). This situation makes it difficult to evaluate directly the magnitudes of jumps and their change with magnetic field.

To analyze the evolution of the first order volume (or length) discontinuity in MnSi with applied magnetic field, we subtract a background contribution. Noting that the dependence of  $\Delta L/L_0 (T)$ in the paramagnetic phase of MnSi in the vicinity of its phase transition can approximated by  linear functions, we subtract these functions from the original data.

The result is illustrated by Fig.~\ref{fig4}. As is seen there, the magnitude of the slightly broadened length discontinuity decreases with applied magnetic field, and at about $\sim 0.26$~T the form of the reduced thermal expansion curves changes qualitatively (Fig.~\ref{fig4}). The discontinuities are replaced by broad anomalies. These anomalies can be still seen at about 0.43 T as evidenced in Fig.~\ref{fig5}, where the thermal expansivity of MnSi is displayed in a limited range of magnetic fields.

The decaying anomaly obviously associated with the one under discussion can be even traced to 0.6~T from the ultrasound studies~\cite{9}.

The sample length discontinuities as function of magnetic field, determined from data in Fig.~\ref{fig4}, are shown in Figs.~\ref{fig6} a and b. It is noteworthy that the discontinuity variations are closely correlated with peak amplitudes of thermal expansivity (Fig.~\ref{fig5}). Some remarkable features in Fig.~\ref{fig6} should be pointed out.

First, in a configuration where the magnetic field is directed along [110] and the linear expansion is measured along [100], $\Delta L/L_0$ initially grows; whereas, in the parallel configuration ($H$, $\Delta L \parallel[110]$) $\Delta L/L_0$ probably decreases from beginning. This situation obviously is connected with a specifics of the longitudinal and transverse magneto-striction in MnSi at low magnetic fields (see      Fig.~\ref{fig3}). Second, a plateau in the range $\sim 0.12-–0.2$~T  (Fig.~\ref{fig6}a) and a local anomaly at $\sim 0.14–-0.23$~T (Fig.~\ref{fig6}b) are related to the existence of the skyrmion A-phase. Third, the sample length discontinuity seemingly drops to zero from  the finite value($\sim 5\cdot 10^{-7}$) at $\sim 0.26$~T.  The forced magnetostriction measurements (Figs.~\ref{fig2},~\ref{fig3}) confirm a lack of first order features at the conical to induced ferromagnetic phase transition on isotherms at 28.6 K and below. A sharp change of slope in the magnetostriction curves identifies a transition of the conical spin structure to field-induced ferromagnetic spin order~\cite{19}.  This transition is associated with tiny dips, which can be seen in Fig.~\ref{fig3}. At the same time, a first order phase transition on the boundary of skyrmion formation is clearly seen (Fig.~\ref{fig3}).

A first order phase transition with a finite jump in thermodynamic quantities across a phase border cannot be terminated without continuation. Normally, a first order phase transition may end in a tricritical point with $\Delta V = 0$ and $\Delta S = 0$. Then, a phase transition may continue as a second order transition; however, finite values of  $\Delta V$, and hence $\Delta S$, at the termination point, a lack of divergence of thermal expansivity $1/L(dL/dT)$ and an unusual deterioration of the related peaks, which should indicate a second order transition, prevent us from accepting this scenario~\cite{17}.  Instead, we hypothesize that the Frenkel concept of heterophase fluctuations may be applicable~\cite{20,21}. Heterophase fluctuations, similar to quenched impurities, smear a first order phase transition starting at some critical value of magnetic field. Consequently, one may expect an enhancement of these fluctuations and deterioration of peaks in thermal expansivity, heat capacity, etc. with magnetic field. The smeared phase transition itself, however, would still exist as it follows from the magnetostriction measurements (see Figs.~\ref{fig2},~\ref{fig3}).
\section{Conclusion}
Thermal expansion and forced magnetostriction of MnSi were measured as a function of temperature to 5~K and magnetic field to 3~T. A small length (volume) discontinuity at the magnetic phase transition in MnSi decreases with application of magnetic field to a value $\Delta L/L \sim 10^{-7}$, and then suddenly the discontinuity seemingly jumps to zero. Thermal expansivity peaks strongly deteriorate with magnetic field. No specific features identifying a tricritical point were observed.  We propose that the Frenkel concept of heterophase fluctuations~\cite{20} may be relevant in the current case. Therefore, we suggest that the magnetic phase transition in MnSi always remains first order at any temperature and magnetic field (see also~\cite{9}),  but the transition is progressively smoothed by heterophase fluctuations.
These results question the applicability of a model of a fluctuation-induced first order phase transition in MnSi. Probably, a model of coupling of an order parameter with other degrees of freedom is more appropriate (see~\cite{22}).
\section{Acknowledgements}
AEP (measurements, data analysis) and  SMS (data analysis, writing the paper) greatly appreciate financial support   of  the Russian Foundation for Basic Research (grant 15-02-02040), Program of the Physics Department of RAS on Strongly Correlated Electron Systems and Program of the Presidium of RAS on Strongly Compressed Matter. AEP is also grateful to the Russian Science Foundation (14-22-00093) for financial support.
	
\end{document}